\documentclass[12pt,a4paper]{article}
\usepackage[T1]{fontenc}
\usepackage{indentfirst}
\usepackage[english]{babel}
\usepackage{natbib}
\setcitestyle{authoryear,open={(},close={)}}
\usepackage{amsfonts,latexsym}
\usepackage{amsthm}
\usepackage{euscript}
\usepackage{amsmath}
\usepackage{color}
\usepackage{graphicx}
\usepackage{caption}
\usepackage{subcaption}
\numberwithin{equation}{section}
\usepackage{ucs}
\usepackage[utf8x]{inputenc}
\usepackage{amsfonts}
\usepackage{graphicx}
\usepackage{fancybox}
\usepackage{tikz}
\usepackage{rotating}
\usepackage{graphics}
\usepackage{graphicx}
\usepackage{caption}
\usepackage{subcaption}
\usepackage{lineno}

\newfont{\bcb}{msbm10 scaled 1200}
\newfont{\bcc}{msbm10}

\title{Centrality-oriented Causality%Outlyingness-oriented
--A Study of EU Agricultural Subsidies and Digital Developement in Poland}
\author{Daniel Kosiorowski (corresponding author)$^1$, \\ Jerzy P. Rydlewski$^2$}
\begin{document}
\maketitle
\begin{center} 
$^1$\textit{Cracow University of Economics, Department of Statistics, Poland, daniel.kosiorowski@uek.krakow.pl}
\\ $^2$\textit{AGH University of Science and Technology, Faculty of Applied Mathematics, Krakow, Poland, \\ ry@agh.edu.pl} 

\end{center}

\begin{abstract}
Results of a convincing causal statistical inference related to socio-economic phenomena are  treated as especially desired background for conducting various socio-economic programs or government interventions. Unfortunately, quite often real socio-economic issues do not fulfill restrictive assumptions of procedures of causal analysis proposed in the literature. This paper indicates certain empirical challenges and conceptual opportunities related to applications of procedures of data depth concept into a process of causal inference as to socio-economic phenomena. We show, how to apply a statistical functional depths in order to indicate factual and counterfactual distributions commonly used within procedures of causal inference. Thus a modification of Rubin causality concept is proposed, i.e., a centrality-oriented causality concept. The presented framework is especially useful in a context of conducting causal inference basing on official statistics, i.e., basing on already existing databases.\\ Methodological considerations related to extremal depth, modified band depth, Fraiman-Muniz depth, and multivariate Wilcoxon sum rank statistic are illustrated by means of example related to a study of an impact of EU direct agricultural subsidies on a digital development in Poland in a period of 2012-2019. 
\end{abstract}

\section{Introduction}
\label{S:1}
In a history of scientific and in particular economic thought one may find various statistical frameworks, which have been developed for expressing necessary assumptions under which statistical results can be endowed with causal interpretation. There is an agreement, that causes should tell us not only that two phenomena are related, but how they are related. They allow us to make reliable prediction about the future, explain the relationships between and occurrence of events and enable to develop effective policies intervention. A causal knowledge should exhibit a certain kind of stability, resistance with respect to possible circumstances under which a phenomenon is observed and with respect to parameters of environment. Causality alone is insufficient basis for undertaking further scientific or decisive acts. In order to use causes to effectively predict, explain or alter behaviour, we must also know the time over which a relationship takes place, the probability with which it will occur, and how other factors alter its efficacy. We need to know when the effect starts and how long it persists. Causal relationships depend on time scales as relationships may persist a while or a long time.
\\ In economics, for several merit justified reasons, causal relations, which are  characteristic for an influential majority of objects or agents are of a prime importance (see \cite{Welfare}). An effective way of defining this influential majority of economic objects provide statistical depth function (see \cite{Zuo:2000}), which is a basic notion used within a discipline of multivariate statistics called \emph{data depth concept} (DDC). Using it one may propose very useful multivariate (see \cite{LPS}, \cite{Liu:1995}) or functional (\cite{Fraiman2001,LopezRomo,Nagy2017,Nair}) generalizations  of one-dimensional statistical procedures based on order statistics and ranks and in a consequence robust causal inferential procedures. Generally speaking, statistical depth function provides a center-outward order of multivariate or functional objects. It expresses a centrality of an object as a number from an interval $[0,1]$, with values close to $0$ treated as peripheral and values close to $1$ as central. Thus one may consider causal relationships of objects, for which measures of departure from centers are not bigger then prefixed thresholds. This is a natural way of focusing our attention on the influential majority of economic objects.
\\ DDC provides a variety of useful statistical procedures (see \cite{Kos1}) but despite excellent findigs obtained in recent years ( \cite{LiuZuoWang2013}, \cite{DyckMoz2016}) its procedures still exhibit significant limitations in a context of conducting empirical research basing on relatively small, sparse  data-sets drawn from resources from statistical offices, with observations often exhibiting "malicious configurations".
\vskip 1mm
One of aims of this paper is to critically discuss selected chances and limitations related to applications of certain tools of DDC in case of robust empirical causal analysis concerning two spatial economic phenomena, namely agricultural subsidies (treatment) and digital development of a country (effect). In the considered empirical example there is a functional dataset as an input set, and a multivariate dataset describing digital development as an output set. Causal inference scheme should be adjusted to these datasets. As it is virtually impossible to repeat the experiment, we treat the typical observations as factual and atypical observations as counterfactual.
\\ Note, that knowledge of such "centre focused causality" seems to be especially desirable in a context of performing social programs dedicated to typical objects (e.g., middle class stimulating program) or for decreasing a fraction of certain atypical objects (e.g., avoiding poverty program). 
\vskip 1mm
In statistical literature, one of the most widely recognized conceptualization of causal inference is Rubin's potential outcome representation (see  \cite{Rubin1974,Dawid:2000,CoxWermuth}). Rubin postulates, that causal statements can only be derived if one additionally considers what would have happened if an object had experienced something different than its experience. Another especially influential in economic and econometric literature concept is Granger non-causality (GNC, \cite{Hendry2017}) and its variants or generalizations. Its empirical version in essence may be expressed as a prediction error approach. Assuming a specific implementation of the GNC one can easily indicate depth based estimation or testing procedure replacing least squares or maximal likelihood principles used by default. Next influential approach to causality analysis are probabilistic approaches (see \cite{Pearl, Kleinberg}). The basic probabilistic theory of causality is that $C$ is a cause of $E$ if 
\begin{equation}
    P(E|C)>P(E|\sim C),
    \label{eq1}
\end{equation}
where $P(E|C)$ denotes conditional probability of E under condition C, $"\sim "$ denotes complement of an event. 
\\ Appealing to the DDC it is straightforward to consider this condition within a central region of a certain degree, and hence focusing on a certain majority of objects. 

\section{Applying DDC in the causal inference}
Nowadays for accepting or rejecting an existence of a certain causal relationship between phenomena, one needs to present convincing empirical evidence obtained in a process of statistical inference. Generally, in order to identify a phenomenon X as a cause of a phenomenon Y one has to demonstrate that a variability in X produce a variability in Y. In this context one may consider various kinds of variations: across time, across individuals, across characteristics, across groups and across intervention versus observation (see \cite{CoxWermuth,Hendry2017}).
\\  Currently it seems to be more and more commonly recognized, that properties of the statistical inference strongly depend on a quality of empirical data used in the inference as well as on fulfilling assumptions underlying the procedures (\cite{Wilcox}).
Using depths we, among others, can demonstrate how robustly measured variability of X produce robustly measured  variability of Y, where X and Y are expressed as multivariate of functional random variables. 
\\ In the spirit of \cite{Rubin1974} the causal effect of one treatment $E$ over another $C$ for  a particular unit and time interval from $t_1$ to $t_2$ is a difference between what would have happened at time $t_2$ if the unit had been exposed to E initiated at $t_1$ and what would have happened at time $t_2$ if the unit had been exposed to $C$ initiated at $t_1$.
\\ \cite{Rubin1974} considered $2N$ units (e.g. small regions of a country), half being exposed to a control treatment (C) and half to a treatment (E). If treatments E and C were assigned to units randomly, $y(E)$ denotes value o $Y$ measured at $t_2$ for the unit given the unit received experimental treatment $E$ initiated at $t_1$, similarly $Y(C)$, then $Y(E)-Y(C)$ is causal effect. \\
The problem in measuring $Y(E)-Y(C)$ is that we can never observe both $Y(E)$ and $Y(C)$ since we cannot return to time $t_1$ to give other treatment. Next problematic issue arises, as in practice an usage of truly random samples may be very difficult (see \cite{Kleinberg}).
\vskip 1mm
Although robustness and causality conceptually seem to be very closely tied - surprisingly considerations binding them are relatively rare. From other point of view, one may get an impression that authors conducting researches on causality in the area of economics in fact implicitly assume some kind of "sample-to-sample stability" of their empirical argumentation even if in a center of their statistical considerations stands evidently non-robust least squares method (see \cite{EngleWhite}). For proposing truly robust causality analysis, it is natural to source from concepts and  developments of modern robust statistics (see \cite{Wilcox}). For several merit-justified reasons related to the fact that economic systems are multidimensional by default, we focused our attention on the DDC methods.
\\ There are many excellent papers presenting particular DDC tools, which are of minor importance for the aims of this paper. 
Here we would like to stress opportunities related to applications of depths for functional objects (e.g., "subsidies trajectories", see \cite{Nieto}) and depth-induced multivariate rank tests (e.g., a comparison of control and treatment groups in terms of various multivariate rank tests basing on ranks induced by depths (see \cite{Jure:2012,Liu:1995,StatPap}). A choice of a specific depth in this context is a statistical as well as merit issue, as the DDC offers a rich overview of depths in terms of a balance between effectiveness, robustness and computational complexivity (see  \cite{LiuZuoWang2013,DyckMoz2016}) and thanks to a locality concept (see \cite{Pain:2013}) options as to "resolution" at which we compare phenomena.  
Moreover, instead of using the propensity scores (see  \cite{Rosenbaum_Rubin1983}) one may consider conducting a family of local Wilcoxon tests for a certain sequence of locality parameters $\beta_1,...,\beta_k$ (see \cite{StatPap}). In the case of a significant difference we expect that majority of the tests reject equality of distributions in the control and intervention group.

\subsection{Depth-based outlyingness and causality}
Formula (\ref{eq1}) gives us some causality inference in a rather simple setup, where $C$ or $\sim C$ occurs.
If the probability space is more complex, (\ref{eq1}) should be a base for some more complex inference.
We are interested in assessing whether the event $\textbf{C}=\{C$ \textit{is one of the central observations in the considered dataset with the locality parameter} $\beta$ $\}$ is a cause of $A$, where locality parameter $\beta$ is understood in the sense of \cite{Pain:2013}, and hence it is a parameter combining probability and outlyingness. For clarity, note that $\sim \textbf{C}=\{C$ \textit{is not one of the central observations in the considered dataset with the locality parameter} $\beta$ $\}$.
\\ Look at the random variable, which is the difference
\begin{equation}
E(A|\textbf{C})-E(A|\sim\textbf{C}).
\label{eq2}
\end{equation}
Realization of the above random variable is a function. 
If 
$$\mu \{x: E(A(x)|\textbf{C})-E(A(x)|\sim\textbf{C})\}=\mu(D)$$
then $C$ is a cause of $A$, where $\mu(D)$ is a Lebesgue measure of the domain $D$ of the random function $A$.
\\ We can choose a level $\beta$, e.g., it can be equal to $99\%$, $98\%$ or $95\%$.
It seems reasonable to treat the quantity  
$$\mu \{x: E(A(x)|\textbf{C})-E(A(x)|\sim\textbf{C})\}=\beta\cdot \mu(D)$$
as a measure of strength of causal relationship restricted to majority of objects of level $\beta$.
On a theoretical level we use a notion of conditional expected value, which is often quite difficult to calculate and estimate for functional objects (see \cite{Bosq}). As robustness stands in the center of our considerations, we propose to use a difference between sample functional medians as estimator of the above quantity.
\\ DDC offers robust measures of multivariate location and scatter, robust regression as well as robust goodness of fit, prediction error measures, and hence it provides natural statistical tools of causal inference. 
\\ Let $\mathbf{Y}=\{Y_1,...,Y_k\}$ denote k-dimensional vector of our interest. The DDC enables, among others, to consider the following causal reasoning schemes:
\begin{enumerate}
    \item{Assessing the difference $\mathbf{Y(E)}-\mathbf{Y(C)}$ using depths (see (\ref{eq1})).}
    \item{For two samples of multivariate or functional objects $\mathbf{Y^n(E)}$ and $\mathbf{Y^m(C)}$ consider a difference between two depth induced sample medians \\ $MED[\mathbf{Y^n(E)}]-MED[\mathbf{Y^m(C)}]$.
    \\ One may expect, that estimated distribution of a difference between two depth induced medians should be informative in a context of causal inference.}
    \item{For two multivariate or functional "regression samples" $\mathbf{Z^n(E)}$ and $\mathbf{Z^m(C)}$ consider distribution of a difference between two vectors of regression parameters $\Theta$ estimated via depth based procedure, e.g., using depth trimmed samples we would like to estimate the distribution $\Theta[\mathbf{Y^n(E)}]-\Theta[\mathbf{Y^m(C)}]$. The distribution should inform us about details of the treatment effect. }
    \item{Depths enable for comparisons of control and treatment groups on sets $C(\beta_i)\backslash C(\beta_j)$, where $C(\beta_i)\backslash C(\beta_j)$ denotes set subtraction of central regions $C(\beta_i),C(\beta_j)$, and $\beta_i,\beta_j$ ($\beta_i>\beta_j$) denote both level of centrality and locality levels (i.e., central regions covering  the smallest depth region with probability equal or larger than $\beta$ in the sense of \cite{Pain}), what can be interpreted as a sort of a comparison on levels appealing to the comparison using propensity scores of \cite{Rosenbaum_Rubin1983}.}

%\item{The DDC enables for a natural implementation of GNC via generalized exponential smoothing (\cite{SIT}).}
\end{enumerate}
In order to effectively use the above approaches in the causal inference we need operationally feasible theory for the listed tools. Unfortunately, with several exceptions depth based inference procedures are conceptually more demanding than classical ones and need asymptotic or re-sampling machinery. The issue is especially evident in the case of a theory of the DDC tools dedicated for functional data (see \cite{Bongiorno1,Bongiorno2}).  

\subsection{A framework for causality analysis using functional and multivariate depths}
Without doubts our perception of an economic phenomenon often relates to an evaluation of properties of a function of a certain continuum. One may consider probability density function of random variable describing an income of a household, one may consider GDP per capita trajectory of a country during a decade, day and night number of visits of Internet users in the Internet service, or a behaviour of an investor optimism indicator within a month. Reducing the whole function to a certain set of scalars (e.g., mean, variance) very often implies a significant loss of valuable information on the phenomenon and in a consequence may lead to inappropriate perception of the phenomenon. A "shape" of the consumer price index (CPI) during a month may better express an investor optimism during the considered period as a specific sequence of peaks and valleys in CPI trajectory may denote sequence of an activity bursts and consumer hesitations, hence "a spectrum of moods" called optimism. Considering economic phenomena as functions is natural and enables us to take into account more information than in “classical scalar approach”. One may associate values of economic aggregates with shapes and other properties of other functional aggregates (e.g. yield curves, Lorenz curves, fertility curves, life expectancy curves etc.). 
\subsection{Proposal}
\label{Proposal}
We have $V$ objects, for which characteristics $\{S^v(t)\}_{v=1,...,V}$ are given for $t\in (t_0, t_1)$, and thus are treated as functional objects. Moreover, a multidimensional vector representing treatment effect for each object is given, namely $A^v\in \mathbf{R}^l$.
Centrality of an object $v_0$ is expressed in terms of value of empirical functional depth for $S^{v_0}(t)$ among $\{S^v(t)\}_{v=1,...,V}$.
Then we divide original $V$ objects into two groups according to higher ($v\in F$), or lower ($v \in C$) value of calculated functional depth. 
Subsequently, rank is calculated for each vector $A^v$ with respect to a chosen multidimensional depth.
In the last step distribution of ranks for both groups, i.e. $A^v$ where $v\in F$ or $v \in C$, is compared using Wilcoxon rank-sum test. 
This centrality-oriented causality is a modification of Rubin causality concept, in this sense that it enables to use a DDC and its tools to define factuals and counterfactuals.

\vskip 2mm Note that similarly we are able to define an outlyingness-oriented causality. Namely we divide original $V$ objects into two groups. The first group ($v\in F$) consist of observations with low value of calculated functional depth, which at the same time are frequently much above (or below) a functional median. The second group ($v \in C$) consist of the remaining observations, or some chosen subset of them.

\vskip 2mm Note also that it seems reasonable to fix outlyingness parameter defining groups $C$ and $F$ so that the number of observations in group $C$ is significantly smaller than the number of observations in group $F$. The outlyingness parameter should depend on the considered depth.
\section{Empirical application}
\begin{figure*}
\includegraphics[width=\textwidth]{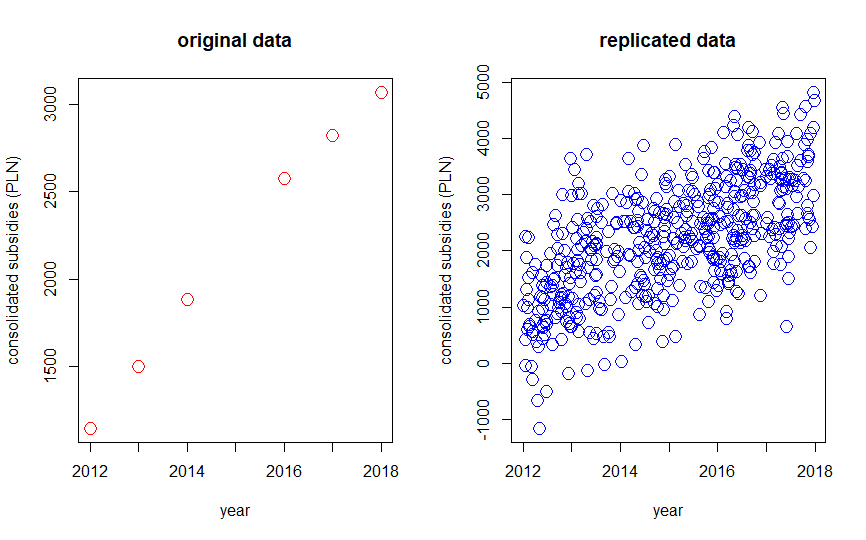}
\caption{Original and replicated data for Opolskie voivodship.}
\label{fig:1}
\end{figure*}

%In \cite{Kosiorowska} certain data-analytic applications of depth based framework to evaluation of the fourth millennium goal has been proposed. A discrimination between statistical, causal and data-analytic inference has not been done however.\\
Main aim of a below example is to underline certain difficulties appearing in real applications of DDC based procedures used in robust causal inference as to phenomena responsible for a development of a country consisted of smaller units, e.g., voivodships or counties. These issues are of prime importance for designing various government programs and evaluating their effects. Generally speaking, conditions we face in practice of economic reasoning significantly depart from the conditions postulated in a theory of causal inference (see \cite{Rubin2005,Dawid:2000}).
\\ We consider an issue of an impact of selected agricultural subsidies granted to farmers in Polish voivodships within European Union (EU) funds (we have used data on amount of direct payment for Campaign 2007-2017 in single area payment scheme; subsidies involved single area payment, subsidies to legumes, subsidies to tomatoes, subsidies to soft fruits)
on a digital development of Poland in a period of 2012-2019 (measured via selected indicators of information society, innovation activity and a degree of dissemination of a telecommunication net). Methods used within our considerations may easily be broaden to similar issues.
In other words, in order to show an application of the proposed approach we have considered a problem of evaluation of an impact of EU agricultural subsidies (an indirect intervention/treatment) for a certain kind of regional development of a country namely digital development of Poland.\\
In "the closest to ideal" circumstances we could observe a representative random sample drawn from all the smallest sub-regions of Poland (i.e., 2477 communities) with respect to an objective measure of digital development at time $t_0$, when a subsidy is taken (in a situation of  
being exposed to a treatment) and at certain time $t_1>t_0$, and could observe the same collection of communities with respect to the same measure of digital development at moments of time 
$t_0, t_1$ when subsidies are not taken (control group). \vskip1mm
In practice, however, basing on Statistics Poland (see \cite{gus}) services we dispose a pair of short already realized time series of treatment variables and response variables concerning bigger sub-regions of higher administrative level of a country  (i.e., 16 Polish voivodships).
In a general spirit of an approach of \cite{GillRobins} we decided to propose the following robust causal inference scheme. For several organizational and ethical reasons our degree of control over the "experiment" is very low. As very often in the practice of economic studies we are given data, and then we are looking for an appropriate model for them. In the causal inference we additionally have to take into account a discrimination between a factual (what happens due to a treatment) and the counterfactual (what would have happen if an alternative treatment had been taken) distribution of a variable expressing an effect of a treatment. One of main merit tasks in this context is to indicate reliable alternative for the treatment.
\\ In the considered empirical example another problems appear. Firstly, number of objects is relatively small. Secondly, number of observations per unit is very small. Namely, we are given a data consisting of $4$ types of agricultural subsidies obtained by each of $16$  Polish voivodships (regions) every year in a period of $2012$ to $2019$ and we consider $6$ available variables representing the digital development of a region.
\vskip1mm
According to the Proposal described in subsection \ref{Proposal}, in the first step we define a certain synthetic variable $S^v=h(S^v_1,...,S^v_{4})$ representing all $4$ types of agricultural subsidies and calculate its value for each voivodship, where $h$ in our example is a sum of all subsidies divided by a population in the year. We treat $S^v$ as a functional object, and hence we have $16$ trajectories of the functional variable $S$. Furthermore, for each voivodship $v=1,..,16$ we have a vector variable $A^{v}=(A_1,...,A_6)$, where $A_i\in \mathbf{R}$, where coordinates of the vector represent measures of certain aspects of digital development of the voivodship. Note that components of the aggregate $A^{v}$ are incomparable. In our case they are expenses expressed in currency (Polish Zloty, PLN), number of some goods or fraction of companies fulfilling certain conditions connected with digital development. In this context an application of multidimensional rank tests seems to be a reasonable solution. Following our general conceptual scheme we replace the multidimensional observations for each voivodship by their ranks induced by multivariate depth (in the example we have used the projection depth calculated via exact algorithm implemented in \cite{ProjMatlab2015}).
\\ In the second step we calculate functional depth for $S^v$, $v=1,...,16$. We use modified band depth, Fraiman-Muniz depth and extremal depth for the purpose. 
As we consider a centrality-oriented causality scheme, we divide voivodships into two groups (subsets of indices $F,C$) according to higher ($v\in F$) and lower ($v \in C$) value of calculated functional depth. In our simulation study regarding properties of the extremal depth we use a depth value of $\alpha=0.5$ to separate sets $C$ and $F$.
\\ In the third step we calculate multivariate Wilcoxon sum rank statistic for the two samples $A^v$, $v\in F$ or $v \in C$.
\\ Note, that for calculating functional empirical depth we need sufficiently larger number of observation per unit (voivodship) than number of units (voivodships). This requirement is sometimes difficult to fulfill, so we need to apply a certain "replication of data" strategy. Without the strategy in our empirical example we cannot use the extremal depth, which possess relevant statistical properties with regard to coverage properties of depth-induced central regions.
\vskip1mm
Due to a fact, that in the considered empirical example increasing linear trends for each voivodsip are quite evident, we decided to substitute the original data-set by simulated artificial data-set consisting of significantly more observations per voivodship than the original data by generating observations from a mechanism, which does not change the qualitative properties of the whole sample. In our example for this reason we have used a simple linear model with parameters estimated via deepest regression method (see \cite{Hubert:1999}) and error characteristics estimated from original data with assumption of normality (see Figure \ref{fig:1}). 
\\ Figure \ref{fig:1} presents an original sequence of consolidated subsidies in a period of 2012-2018 (a data for 2015 is missing in Statistics Poland database) for Opolskie voivodship (left panel) and a simulated sequence of 500 observations (right panel) from a model "imitating" the left panel data and based on a simple linear trend estimated via the deepest regression method and error characteristics estimated from the original data with normality assumption.

In order to treat the treatment as a cause of the aggregate $A$ we have to show that a conditional distribution of $A$ under condition of the factual subsidies differs from a conditional distribution of $A$ under certain alternative treatment plan -- an alternative sequence of subsidies. A fundamental difficulty here is that we have only an access to realized database provided by the Statistics Poland (https://stat.gov.pl/ \cite{gus}) and we obviously do not observe counter factual distribution, i.e., distribution of $A$ under different sequence of subsidies.
\vskip 1mm
In order to overcome this difficulty and taking into account a postulate stating that causal inference should possess a certain stability property we propose to compare a distribution of the aggregate $A$ estimated basing on voivodships with indices $v\in F$, which are central (i.e., typical) in terms of subsidies trajectory with a distribution of the aggregate $A$ estimated basing on voivodships with indices $v\in C$, which are peripheral in terms of subsidies trajectory. The first distribution is treated as factual whereas the second distribution as counteractual.
\\ Note that we have a functional dataset as an input set, and a multivariate dataset as an output set. We have to adapt a causal inference scheme to these datasets.
The proposed approach is arguable, but note that we assume that there exists a feasible digital development path for the considered homogeneous units, i.e., the regional digital development path can be achieved in the real world. 
 Moreover, it is virtually impossible to repeat the experiment. That is why we treat the typical or central observations as factual and atypical observations as counterfactual.
%\\ (*) ZDANIE DLACZEGO: na wejsciu dane funkcjonalne, na wyjsciu multivariate, dodatkowo nie da sie powtórzyć eksperymentu, wiec trzeba zmodyfikowac procedure wniskowania przyczynowego. to podejscie moze byc dyskusyjne, ale zakladamy ze istnieje dostepna (ktora moze byc zrealizowana) ścieżka typowego rozwoju cyfrowego. Ponieważ zbiór danych/pomiar jaki mamy nie moze byc w żaden sposób powtórzony, to to co mamy typowego jest factual, a nietypowego counterfactual.
In other words, we compare the digital development aggregate $A$ conditioned on a treatment identified with subsidies trajectories with high degree of centrality with aggregate $A$ conditioned on subsidies trajectories with a low degree of centrality.
\\ In practice, first we divide voivodships with respect to specified levels of centrality, i.e., probability coverage, and then we estimate appropriate conditional distributions of aggregate $A$ using certain kind of bootstrap method.
\\ As we use random data replication procedure, we propose to repeat the second step 1000 times for 1000 artificially replicated data-sets and infer on differences between factual and counterfactual distributions basing on estimated distribution of multivariate Wilcoxon sum rank test applied to groups of objects indicated in the second step of the procedure (see \cite{StatPap}). 

\begin{figure*}
\includegraphics[width=0.87\textwidth]{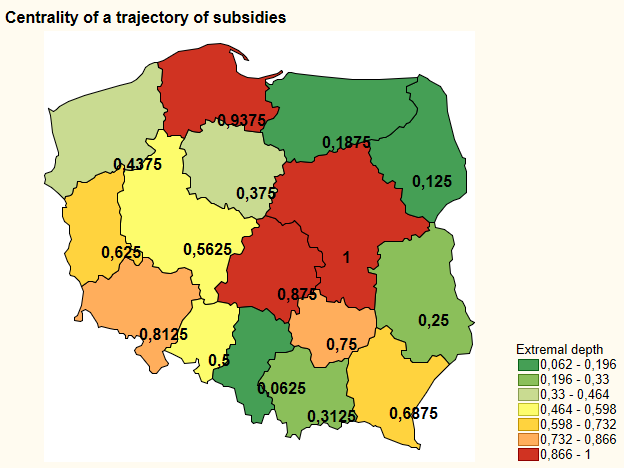}
\caption{Typicality of a trajectory of EU subsidies of Polish voivodships with respect to the extremal depth in 2012-2019.}
\label{fig:2}
\end{figure*}

For measuring the centrality of the subsidies trajectory we use a modified band depth (MBD, \cite{LopezRomo}), Fraiman-Muniz depth (FM, \cite{Fraiman2001}) and extremal depth (ED, \cite{Nair}).
\\ In the considered empirical example we have repeated 100 times a whole sequence of of the second and the third steps repeated 1000 times and we have obtained averages of 1000 average values of the multivariate Wilcoxon sum rank statistic equal to $37.72(3.46)$ when ED has been used, $33.23(0.084)$ when FM has been used and $24.212(0.0734)$ when MBD has been used, where in brackets standard deviations of the 100 means are given (see Table 1, the first row). 
\\ Additionally, in order to strengthen conclusions drawn from our causal inference procedure, we repeated 1000 times an experiment, in which we have compared two independent random samples $A^j$ and $A^k$, where $j$ is the same as number of observations in $F$ and $k$ is the same as number of observations in $C$ and we have obtained an average value of the multivariate Wilcoxon sum rank statistic equal to $32.281(5.960)$ when ED has been used, $31.761(6.69)$ when FM has been used and $29.2692(6.49)$ when MBD has been used (see Table \ref{tab2}, the second row). A significant difference between average Wilcoxon sum rank statistic for samples representing factual and counterfactual distributions chosen via the proposed functional outlyingness criterion and those chosen at random justifies a validity of the proposal. 
\begin{table}
\centering
\begin{tabular}{c||c|c|c}\hline
Type of samples& MBD &FM&ED\\\hline
Sample defined by outlyingness& 24.212(0.0734)&33.23(0.084)	&	37.72(3.46)\\
Sample defined at random& 29.2692(6.49)&31.761(6.69)	&32.281(5.960)	
\end{tabular}
\caption{Comparison of average values (and standard deviations in brackets) of the multivariate Wilcoxon sum rank statistic for factuals and counterfactuals defined by outlyingness (row 1) and factuals and counterfactuals defined completely at random (row 2).}
\label{tab2}
\end{table}
\\ To sum up, MBD behaves better than ED and FM in this sense that it allows for better discrimination between factual and counterfactual distribution, and hence MBD gives the strongest arguments that EU subsidies influence the digital developement of Poland.
\\ In Table \ref{tab1} a comparison of depth values for 16 Polish voivodships is presented. No replication strategy has been applied in MBD and FM case, so only the original data from Statistics Poland has been used. As previously indicated and implicitly stated in \cite{Nair}, in ED case we had to replicate the original data. We conclude therefore from our studies, that agricultural subsidies may be treated as one of causes of digital development in the regions (voivodships) of Poland.
\\ Figure \ref{fig:2} presents an administrative map of Poland with colors representing an empirical extremal depth of the consolidated subsidies trajectory of the voivodship. The figure has been created using ggplot2 and extdepth R packages (see \cite{R,ggplot2,extdepth}).

\begin{figure*}
\includegraphics[width=0.87\textwidth]{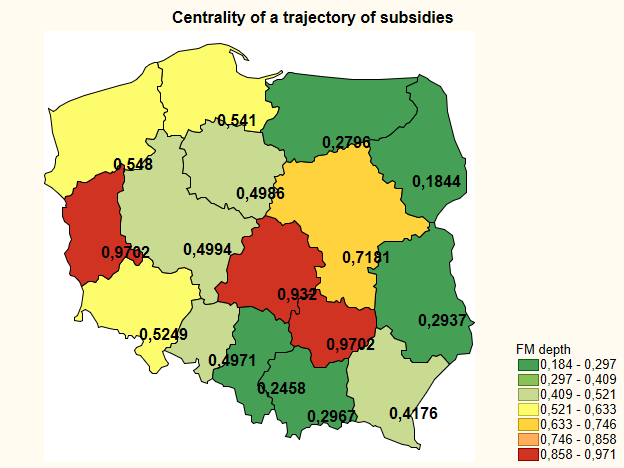}
\caption{Typicality of a trajectory of EU subsidies of Polish voivodships with respect to the FM depth in 2012-2019.}
\label{fig:3}
\end{figure*}

We would like to stress, that we have considered several data replication algorithms including polynomial and constrained polynomial regressions.\\
Although the maximal depth object changed from repetition to repetition, allocations of voivodships within groups $F$ and $C$ exhibited a high level of stability.
\begin{table}
\centering
\begin{tabular}{c||c|c|c}\hline
Voivodship& MBD &FM&ED\\\hline
Dolno\'sl\k{a}skie& 0.425&0.5249	&	0.8125\\
Kujawsko-Pomorskie& 0.5&0.4986	&0.375\\
Lubelskie & 0.341(6)&0.2937	&0.25	\\
Lubuskie&0.591(6)&0.9702&0.625\\
\L \'odzkie &0.575&0.9320&0.875\\
Ma\l opolskie&0.241(6)&0.2967&0.3125\\
Mazowieckie&0.541(6)&0.7181&1\\
Opolskie&0.475&0.4971&0.5\\
Podkarpackie&0.341(6)&0.4179&0.6875\\
Podlaskie&0.125&0.1844&0.125\\
Pomorskie&0.491(6)&0.5410&0.9375\\
\'Sla\k{s}kie&0.125&0.2458&0.0625\\
\'Swi\k{e}tokrzyskie&0.591(6)&0.9702&0.75\\
Warmi\'nsko-Mazurskie&0.241(6)&0.2796&0.1875\\
Wielkopolskie&0.48(3)&0.4994&0.5625\\
Zachodniopomorskie&0.575&0.5480&0.4375
\end{tabular}
\caption{\label{tab:widgets} Comparison of depth values for 16 Polish voivodships.}
\label{tab1}
\end{table}
\section{Summary and conclusions}
Causal inference is commonly treated as an essence of a scientific comprehension of empirical reality. This kind of reasoning has many variants supported by different schools of economic thought. Although the GNC concept seems to be the most popular in the economics, we have focused our attention on less popular in economics but very prominent in other sciences concept proposed by Donald Rubin and modified it in order to obtain a centrality-oriented causality reasoning scheme. As, generally speaking, in economics it is very difficult to conduct a truly randomized experiment postulated by theory of Rubin, we have proposed a certain kind of implementation of his theory, which uses "a trick" based on an application of a depth for functional data and is possible to apply in practice.
\\ In other words we have proposed a novel DDC based method of indicating factual and counterfactual distributions in causal inference scheme of D. Rubin. This seemingly very simple "trick" is of a prime importance in a context of difficulties we face in real empirical causal analysis based on official statistics. \\
We have applied the centrality-oriented causality scheme to study an impact of agricultural subsidies, which may be treated as an indirect intervention or treatment, on a degree of digital development in the regions of Poland in a period of 2012-2019. We have obtained arguments for an existence of causal relation between these  economic phenomena. The proposed scheme may be easily generalized and adjusted to other studies. Having in a disposal richer datasets one can obtain stronger conclusions than these we have obtained.\\    
Thus we have shown, that the DDC offers valuable possibilities for conducting robust causal inference in the economics, especially in multivariate and functional cases. The DDC is not a remedy for solving fundamental issues of causal inference of how to perform causal inference basing on already existing datasets. In order to treat procedures of the DDC as real alternatives to classical multivariate of functional methods further theoretical studies on sample properties of the procedures are required. 
\\ As a result of our studies, we conclude that MBD and FM functional depths are more applicatively useful than ED in a context of analyzing relatively small and sparse empirical data-sets. As ED has other desired statistical properties, further studies on its modifications for relatively small data-sets are required. Original data-set and simple R code, which have been used in the paper are available upon request.
\vskip1mm
\textbf{Acknowledgements} Daniel Kosiorowski thanks for financial support from the Polish Ministry of Science and Higher Education within “Regional Initiative of Excellence” Programme for 2019-2022. Project no.: 021/RID/2018/19. Total financing: 11 897 131,40 PLN.", Daniel Kosiorowski thanks for the support related to CUE grant for the research resources preservation 2019.\\ Jerzy P. Rydlewski research has been partially supported by the AGH UST local grant no. 16.16.420.054. 

\bibliography{literaturaJPR.bib}
\end{document}